\documentclass[10pt]{article}

\usepackage{amsmath}
\usepackage{amssymb}

\usepackage{graphicx}

\usepackage{cite}

\usepackage{color}

\usepackage{multirow}

\usepackage[labelfont=bf,labelsep=period,justification=raggedright]{caption}

\makeatletter
\renewcommand{\@biblabel}[1]{\quad#1.}
\makeatother

\date{}

\begin{document}

\begin{flushleft}
{\Large
\textbf{Efficient Multiple Object Tracking Using Mutually Repulsive Active Membranes}
}
\\
Yi Deng$^{1}$,
Philip Coen$^{2}$,
Mingzhai Sun$^{3}$,
Joshua W.~Shaevitz$^{4,\ast}$
\\
\bf{1} Department of Physics, Princeton University, Princeton, NJ, U.S.A
\\
\bf{2} Department of Molecular Biology, Princeton University, Princeton, NJ, U.S.A
\\
\bf{3} Lewis-Sigler Institute for Integrative Genomics, Princeton University, Princeton, NJ, U.S.A
\\
\bf{4} Department of Physics and Lewis-Sigler Institute for Integrative Genomics, Princeton University, Princeton, NJ, U.S.A
\\
$\ast$ E-mail: Corresponding shaevitz@princeton.edu
\end{flushleft}

\section*{Abstract}
Studies of social and group behavior in interacting organisms require high-throughput analysis of the motion of a large number of individual subjects. Computer vision techniques offer solutions to specific tracking problems, and allow automated and efficient tracking with minimal human intervention.  In this work, we adopt the open active contour model to track the trajectories of moving objects at high density.  We add repulsive interactions between open contours to the original model, treat the trajectories as an extrusion in the temporal dimension, and show applications to two tracking problems.  The walking behavior of \textit{Drosophila} is studied at different population density and gender composition.  We demonstrate that individual male flies have distinct walking signatures, and that the social interaction between flies in a mixed gender arena is gender specific.  We also apply our model to studies of trajectories of gliding \textit{Myxococcus xanthus} bacteria at high density.  We examine the individual gliding behavioral statistics in terms of the gliding speed distribution.  Using these two examples at very distinctive spatial scales, we illustrate the use of our algorithm on tracking both short rigid bodies (\textit{Drosophila}) and long flexible objects (\textit{Myxococcus xanthus}).  Our repulsive active membrane model reaches error rates better than $5\times 10^{-6}$ per fly per second for \textit{Drosophila} tracking and comparable results for \textit{Myxococcus xanthus}.

\section*{Author Summary}
Many living organisms exhibit complex interactions between individuals.  The number of individuals involved in social behaviors ranges from two for the case of a courting pair to thousands or more. Studies of the social behavior from bacteria to animal populations can reveal the underlying mechanisms connecting individual behavior to group phenomena. These types of analyses are often dependent on the ability to track each individual over time.  Computer vision is a powerful tool for identifying objects in images and movies, and  computer-aided video tracking has become a common step in processing raw video data of behaving organisms.  Here, we present a tracking routine that is capable of coping with densely packed objects that are in close contact with one another.  We test our algorithm on movies from two vastly different social model organisms: the gliding bacteria \textit{Myxococcus xanthus} and the fruit fly \textit{Drosophila melanogaster}.  Our algorithm achieves very low error rates and high computational efficiency with minimal human intervention.  

\section*{Introduction}
A broad range of biological problems on many length scales, from cells to  whole animals, require the ability to track moving individuals with a group.
Advances in computer vision in the past two decades have enabled  computed-aided automatic or semi-automatic tracking programs to greatly boost the capacity to analyze large amounts of data and reduce the involvement of human observers.  However, many traditional tracking algorithms struggle when objects come into close physical contact or even overlap within an image.
Recent work  addressed this problem by separating  objects using a Gaussian mixture model with an area prior followed by identity matching and successfully applied this approach to track walking \textit{Drosophila} with minimal human supervision \cite{Branson2009}.  Tsai and Huang further extended this approach by refining the segmentation of the \textit{Drosophila} images  into different body parts which enables more detailed measurements to be made \cite{Tsai2012}.  A non-Bayesian framework was used by Chaumont et al. to track multiple mice by modeling the animal body with a series of physical ``primitives'' connected by joints and elastic springs that can interact with each other \cite{Chaumont2012}.  All of these algorithms are capable of handling large amounts of images, $>10^4$ frames,  with relatively little tracking error that is then corrected manually.  However, this error rate increases with reduced image quality or when the objects move close to one another more frequently.

In most tracking solutions, image segmentation is performed on each frame to identify individual objects. This is then followed by an identity matching scheme between frames based on distance,  object birth and death probabilities, and other estimated parameters \cite{Jaqaman2008, Branson2009}.  Active contours (snakes) are a popular image segmentation approach that is widely applied in analyzing biological and medical images.  The contour of the compartment boundary is treated as an elastic band that interacts with the image and exhibits a damped relaxation to the minimum energy state \cite{Kass1988}.  In addition to closed-contour uses for measuring parameters like object area, open active contours can be used to detect filamentous objects such as blood vessels\cite{Can1999, Kirbas2004, Altinok2011}, neurons\cite{Vasilkoski2009, Meijering2010, Wang2011}, actin filaments\cite{Li2010, Xu2011}, and many biomedical and non-biomedical applications. In these approaches, active contours can be allowed to merge, break, fork and recombine.

In this work, we solve the tracking problem using a deformable membrane model, which is an extension of the active contour model to the temporal dimension.  In order to prevent merging of multiple objects, we add a repulsive interaction between neighboring contours.  We test this method on two practical tracking problems in animal behavior and microbial ecology: walking \textit{Drosophila} and gliding bacteria \textit{Myxococcus xanthus} are tracked at high density with low error rate (< $5\times 10^{-6}$ per fly per second, or $10^{-5}$ per cell per second), and at high efficiency (better than 50 frames per second when tracking 5 flies).  \textit{Drosophila} has become a popular model organism to study neurobiology and animal behavior for its ease of genetic manipulation\cite{Chan2007, Ho2005, Frye2004, Greenspan2000, Robinson2008, Iyengar2012}, and \textit{Myxococcus xanthus} is a gram-negative bacteria that exhibit gliding motility using molecular motors\cite{Mignot2007, Sun2011}, which allow the bacteria to exhibit complex group behaviors\cite{Kuner1982, Monds2009, Xie2011, Zhang2012}.  Using these two tracking problems at very distinctive spatial scales, shapes and morphology, we demonstrate the application of our algorithm on a broad range of problems.  Finally, we discuss the close connection between Bayesian techniques and the energy minimization approach in our active membrane algorithm.

\section{Models}
\subsection{The classical active contour model}
In the classical active contour model\cite{Kass1988}, a feature in an image, usually a line, area boundary or edge, is located by relaxing an elastic contour (snake) that interacts with the image to the contour's minimum energy state.  The energy of the contour consists of the internal elastic energy and the image energy term $E_{im}$ based on the location of the contour in the image, calculated along the contour $\mathbf x(s)$ in $N$ dimensional space, as a function of arc position $s$:
\begin{equation}
E = \int_A^B \left(\frac{1}{2}\left(\alpha |\mathbf x'(s)|^2 + \beta |\mathbf x''(s)|^2\right)  + E_{im}(\mathbf x)\right)ds,
\end{equation}
where the $\alpha$ term penalizes the energy when the contour is deviated from a uniform straight line and the $\beta$ term adds an additional cost to bending. $A$ and $B$ are the termini of an open contour, which we define as 0 and 1, or in case of a closed contour, the integral path is closed.  Minimizing $E$ is equivalent to solving
\begin{equation}\label{eq::dynamics}
\alpha \mathbf x'' - \beta \mathbf x''''-\nabla_{\mathbf x} E_{im}(\mathbf x) = 0,
\end{equation}
which can be written in the discrete form\cite{Kass1988}:
\begin{equation}\label{eq::discretedynamics}
A^*x_{i} + \frac{\partial E_{im}}{\partial  x_{i}}=0, \quad i = 1, 2, ..., M,
\end{equation}
where each $x_{i}$ is an $M$-element vector that defines the contour using $M$ discrete points that are equally spaced along $s$.  $A^*$ is the $M\times M$ circular pentadiagonal  discrete equivalent of the operator $-\alpha d^2/d s^2 + \beta d^4/d s^4$:
\begin{equation}
A^* =\left(
\begin{array}{cccccccc}
a_0+b_0&a_1+b_1&b_2&0&...&0&b_2&a_1+b_1\\
a_1+b_1&a_0+b_0&a_1+b_1&b_2&0&...&0&b_2\\
b_2&a_1+b_1&a_0+b_0&a_1+b_1&b_2&0&...&0\\
&&&...&...&&&\\
0&...&0&b_2&a_1+b_1&a_0+b_0&a_1+b_1&b_2\\
b_2&0&...&0&b_2&a_1+b_1&a_0+b_0&a_1+b_1\\
a_1+b_1&b_2&0&...&0&b_2&a_1+b_1&a_0+b_0\\
\end{array}
\right),
\end{equation}
with $a_0 = 2\alpha$, $a_1 = -\alpha$, $b_0 = 6\beta$, $b_1 = -4\beta$, $b_2 = \beta$.  Equation \ref{eq::discretedynamics} can then be solved iteratively via
\begin{equation}
 x_{i}^\tau = (A^*+\gamma I)^{-1}\left(\gamma  x_{i}^{\tau-1}-\nabla_{i} E_{im}( x_{i}^{\tau-1})\right),
 \label{eq::iterSolution}
\end{equation}
where $\tau$ is the iteration index,  $\gamma$ sets the time step, and $\nabla_{i}$ defines the spatial derivatives $\partial/\partial x_{i}$.  Additional forces can be conveniently incorporated into the model by adding a force term $F_i$, which we will discuss in the following sections:
\begin{equation}
 x_{i}^\tau = (A^*+\gamma I)^{-1}\left(\gamma  x_{i}^{\tau-1}-\nabla_{i} E_{im}( x_{i}^{\tau-1})+F_i\right).
\end{equation}

Numerous variations of this active contour model have been developed for specific image-analysis and tracking problems, including the use of variable stretching and bending stiffness $\alpha$ and $\beta$, sophisticated image potentials $E_{im}$, and the inclusion of additional forces for specific purposes. Here, we build our model aiming to solve the multiple object tracking problem in time-lapsed movies.

\subsection{Open contours}
While a circular pentadiagonal matrix $A^* $ is suitable for dealing with a closed contour, for open contours, the motion of the two tips needs to be considered separately.  We modify the first and last two rows of $A^*$ in equation \ref{eq::discretedynamics} to $A$ for open contours such that the internal force acting on the endpoints is equal to half of the internal force on the nearest neighbor, but in the opposite direction:
\begin{equation}
A =\left(
\begin{array}{cccccc}
-a_1/2&-a_0/2&-a_1/2&0&...&0\\
a_1&a_0&a_1&0&...&0\\
&&...&...&&\\
0&...&0&a_1&a_0&a_1\\
0&...&0&-a_1/2&-a_0/2&-a_1/2\\
\end{array}
\right).
\end{equation}

Without an additional constraint, the tips are left free to interact with the image.  Tip forces can be added to elongate or shorten the contour in order to control the contour length.  In many multiple object time-lapse tracking problems, the characteristic length scale of the tracked objects is known and does not change, even though the position, orientation, and specific shape may alter over time.  In this work, the length of the contours are maintained constant by a harmonic tip-stretching force, $F_{str}$ depending on the length of the contour $l_0$:
\begin{equation}
\label{eq::stretchingF}
F_{str} = -\kappa(l-l_0) \hat{\rho},
\end{equation}
where $\hat{\rho}$ is the tangential direction of the contour at the endpoints pointing outwards,
\begin{equation}
\hat{\rho} =
\begin{cases}
-\frac{d\mathbf x/ds}{|d\mathbf x/ds|}&\colon s = 0\\
\frac{d\mathbf x/ds}{|d\mathbf x/ds|}&\colon s = 1\\
0&\colon \text{otherwise}
\end{cases},
\end{equation}
 $l$ is the length of the contour $l=\int_{0}^{1}\left|d\mathbf x/ds\right|ds$, and $\kappa$ governs the magnitude of the stretching force.

We also set $\beta = 0$ in our two following sample applications because penalizing the contour length adds an effective energetic cost to path curvature.  We find that a non-zero value for beta does not qualitatively change the behavior of the open contours analyses for these two examples.

\subsection{Track contour motion over time}
The kymograph of a contour that moves in $N$-dimensional space over time is a continuous surface in $N+1$ dimensions.  In certain types of problems where the objects' trajectories are less predictive, such as the case of Brownian motion, a global optimization strategy such as the framework to solve the linear assignment problem (LAP) is usually necessary\cite{Jaqaman2008}.  However, in common cases where the trajectories are smooth relative to the object density, local optimization is sufficient and more efficient.  Here, we apply the concept of active contours to the temporal dimension as well as the spatial dimension within each frame, and optimize the localization of the two-dimensional active membrane in the $N+1$-dimensional kymograph.  Different from freely moving contours in all spatial dimensions, the object has one unique location at each time point, and no localization information is carried between frames.  For these reasons, we allow the control points of the active membrane to move within each time slice but not along the temporal axis.  For active membranes, $x_{it}$ is a $M\times T$ matrix that determines the location of the contour at all time, where $M$ is the number of control points along the contour, and $T$ is the number of frames.  We seek the solution of
\begin{equation}
Ax_{it} +x_{it}B^T + \frac{\partial E_{im}}{\partial  x_{it}} - F_{it}=0, \quad i = 1, 2, ..., N, \quad t = 1, 2, ..., T,
\end{equation}
where the $T\times T$ pentadiagonal matrix $B$ effects the derivatives in time and has the same structure as $A$ but transposed.  Similar to the strategy for solving equation \ref{eq::iterSolution}, this equation can be solved iteratively by
\begin{equation}
(A+\gamma I)x_{it} ^\tau+x_{it}^\tau B^T + \nabla_{i} E_{im}( x_{it}^{\tau-1}) - \gamma x_{it}^{\tau-1} - F_{it}=0,
\end{equation},
which is in the form of a Sylvester equation.

In practice, instead of optimizing the active membrane on all frames, we slice the kymograph into overlapping time-blocks in the temporal dimension and then sequentially obtain the optimized solutions in short blocks in accordance with the results from previous blocks in time.  This is a practical simplification as long as the block size is comparable to the persistence time of motion of the objects.  When the temporal projections of the trajectory of one or multiple objects overlap, incorrect initial condition can cause slow convergence or trap the solution at local minima where the registration of traces to image potential minima is swapped.  By solving the problem in a block-wise manner, we avoid incorrect initial placement of the contours, and greatly reduce the computation time using the local initial guess based on previous results.  Because our model is used to process time-lapsed movies, we use the term contour and membrane interchangeably in the following text.

\subsection{Repulsion between multiple contours}
\begin{figure}[!ht]
  \begin{center}
  \includegraphics[width=4in]{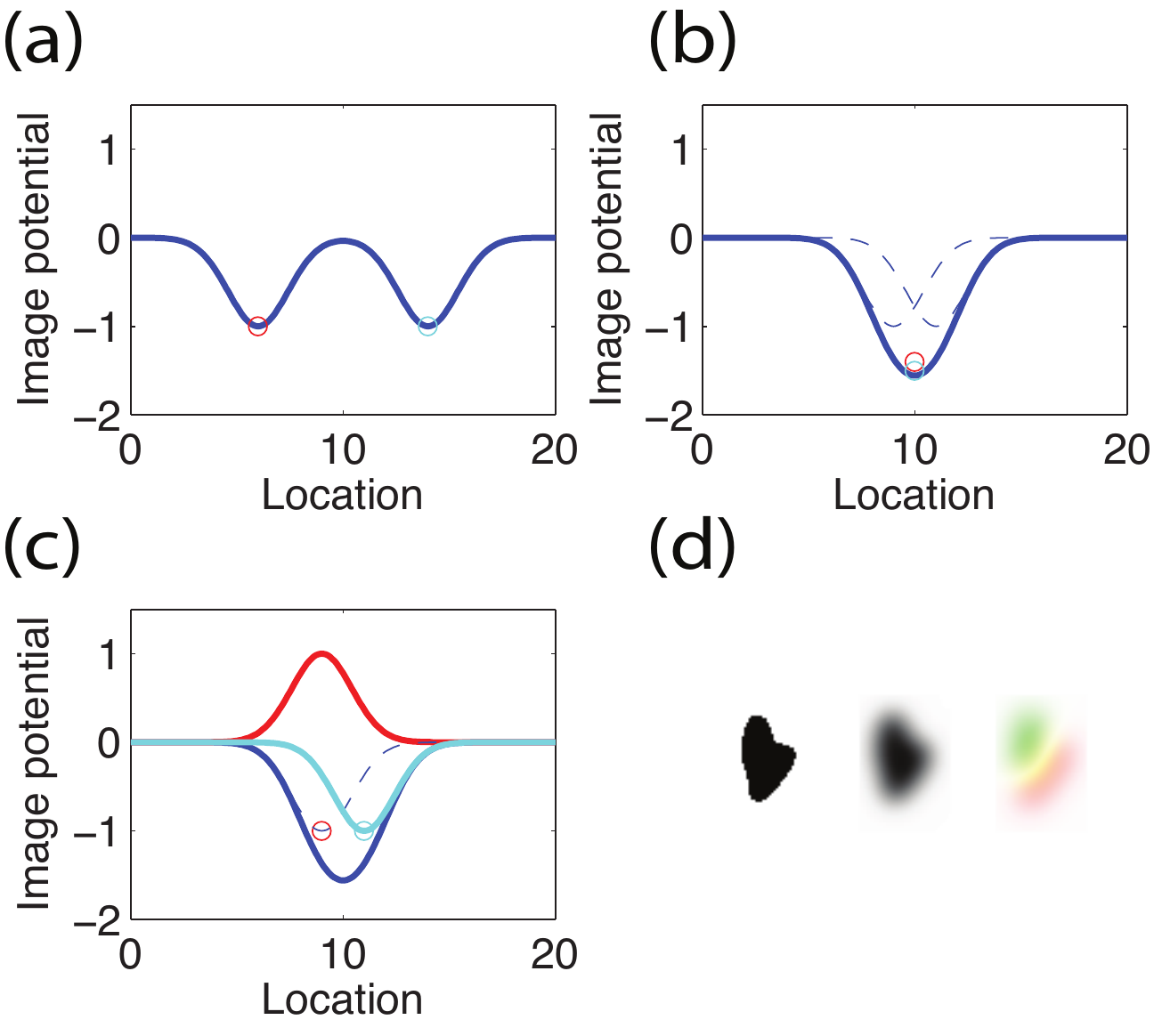}
  \end{center}
  \caption{
  {\bf Schematic illustration of the principle of repelling active contours.} (a) When objects are far away and the attractive image potential fields (blue solid line) don't interact, two active objects (red and cyan circles) correctly fall into the image energy minima.  (b) When objects are close, the potential fields (blue dashed lines) overlap and causes dislocated or merged minima. Two active objects converge into the same minimum. (c) With the repulsive potential added (red solid line), the total field the other object is in is recovered (cyan solid line).  (d) The same principle can be easily applied to two and higher dimensions as shown.  The black solid ovals represent two contacting objects.  The images are blurred to give smooth attraction potential, and with repulsion, two minima can be resolved (green and red).}
  \label{fig::contourcartoon}
\end{figure}

Active contour and surface models can be applied to an image or time-lapse movie of multiple objects by starting at different initial locations.  However, when the objects are in close proximity to one another, the barrier that separates the two objects in the image energy landscape can diminish below a significant level due to noise and other image-based effects (fig. \ref{fig::contourcartoon}).  As a result, two contours (or membranes) with different initial locations can converge to the same image energy minima. To avoid different contours from collapsing into the same pocket, we optimize the energy of multiple contours simultaneously with the addition of a mutual repulsive force between the contours.  Because the repulsive force resembles physical exclusion, we limit the range of the repulsion to the approximate size of the objects, and set the force magnitude to match the depth of the typical depth of the image energy minima.  Specifically, the force is defined to have a quadratic form of the distance between two contours with a cutoff distance $x_0$, which is chosen to be the same as the typical width of the tracked objects:
\begin{equation}\label{eq_repulsion}
  \textbf{F}^{\textrm{repel}}_{lm}(s_l) = \int_{0}^{1} \left[1-\frac{|\mathbf x_l(s_l)-\mathbf x_m(s_m)|}{x_0}\right]\frac{\mathbf x_l(s_l)-\mathbf x_m(s_m)}{x_0}ds_m,
\end{equation}
where $\textbf{F}^{\textrm{repel}}_{lm}(s_l)$ is the unit arc length density of the repulsive force acted on the $l^{th}$ contour at arc position $s_l$ due to the presence of the $m^{th}$ contour.  The discrete version of equation \ref{eq_repulsion} takes the form
\begin{equation}
  \textbf{F}^{\textrm{repel}}_{it;lm} = \sum_{j =1}^N\left[1-\frac{|\mathbf x_{it;l}-\mathbf x_{jt;m}|}{x_0}\right]\frac{\mathbf x_{it;l}-\mathbf x_{jt;m}}{x_0},
\end{equation}
where $\mathbf x_{it;l}$ is the coordinate of the $i^{th}$ control point of the $l^{th}$ contour at the $t^{th}$ frame.  Here, the repulsive force is not normalized to the actual length of the contour, but this is not a significant problem if all contours have similar lengths.  A cubic potential is a sufficiently close approximation to the overall shape of the attractive potential generated by Gaussian smoothing of the original image.

\section{Results}
\subsection{\textit{Drosophila} walking behavior}
We first demonstrate the repulsive, active membrane model on movies of walking fruit flies. \textit{Drosophila melanogaster} has become a popular model system for studying complex behaviors such as courtship, aggression, and learning through the analysis of time-lapse movies of fly position.  In these experiments, individual flies often come physically close to each other causing their images to merge.  The repulsive membrane model is particularly adept at resolving the positions of flies during these events with high chance of contacting and overlapping.

\begin{figure}[!ht]
  \begin{center}
  \includegraphics[width=4in]{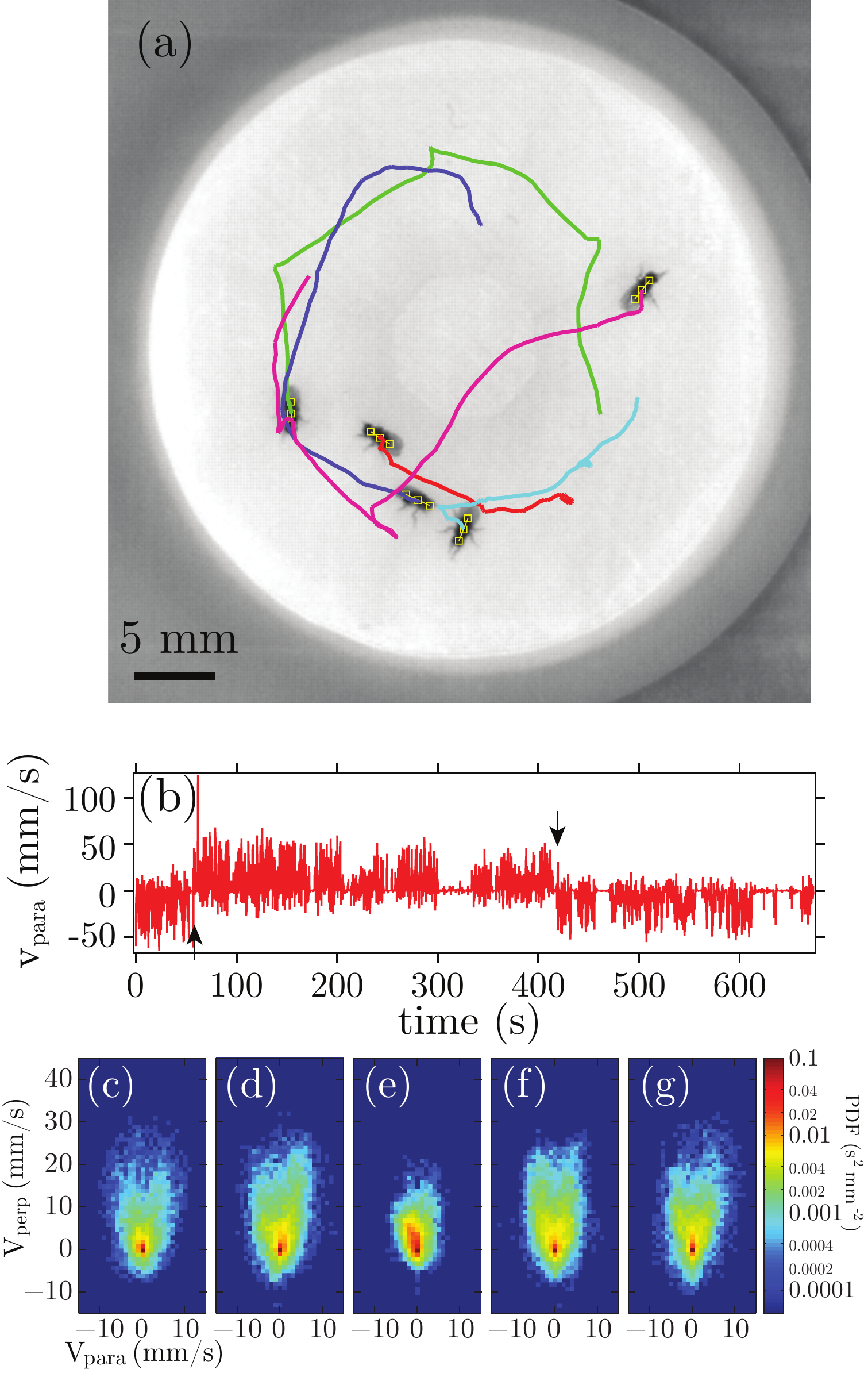}
  \end{center}
  \caption{
  {\bf The trajectories and velocities exacted from the tracking results of five male flies.} (a) A sample image of five flies in the circular arena.  Three seconds of walking trajectories of the flies are labeled in different colors.  The control points are labeled as the yellow squares.  (b) The parallel component of the walking velocity of one fly is plot as a function of time.  When the fly jumps (indicated by arrows), the tracker may reverse the orientation and negate the velocity.  The reversal can be detected off-line using a Hidden Markov Model (HMM). (c-g) Velocity histograms of five individual flies.  The color indicates the the probability density distribution plotted on a logarithmic scale.}
  \label{fig_vStat}
\end{figure}

We recorded a movie of five male flies walking in a circular, 2.5-cm diameter arena at a density of 1 fly/cm$^2$ at 30 frames per second and a camera magnification of 0.1 mm/pixel. Our tracking algorithm models each fly as an open contour with 3 control points (Fig. \ref{fig_vStat}).  The length of each fly is determined from the initial image and remains fixed throughout the tracking task.  Our tracker is able to follow five flies correctly in all 20,203 frames (a total time of 673 seconds).  The oval shaped fly image potential constraints the orientation of the open contour, so that the head direction is correctly resolved in most frames.  Occasionally, a fly can exhibit escape behavior that causes a quick change in location and head direction.  For generality, our algorithm does not consider image details specific for fruit flies, and we do not distinguish head from tail based on images. Consequently, jumps may reverse the head-tail orientation (Fig. \ref{fig_vStat}b). However, because flies mostly move forward in the direction of the head with low jumping frequency on the order of $0.1$~min$^{-1}$, a Hidden Markov method is suitable for detecting jumps.  We use this information and adopt the Viterbi algorithm\cite{Cormen2001} to determine the forward-backward head direction as an addition step of our fly tracking software. Figure \ref{fig_vStat}(c-g) shows the 2D,  $\log$-probability distribution function of walking velocity for each of the five flies after the head direction correction.  Each fly has an individual signature distribution that is slightly different from other's.

\begin{figure}[!ht]
  \begin{center}
  \includegraphics[width=4in]{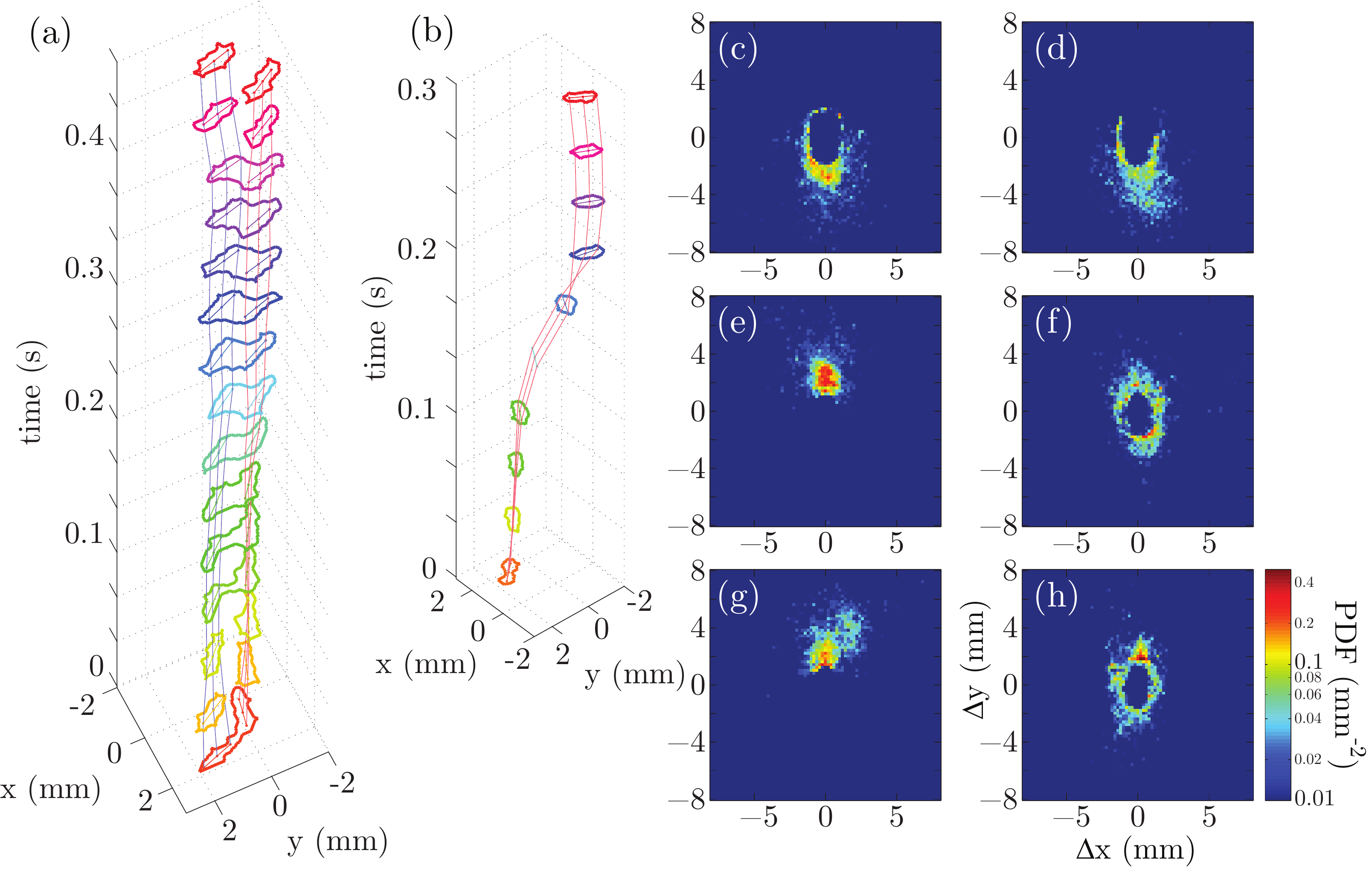}
  \end{center}
  \caption{
  {\bf Tracking results of in-contact and jumping \textit{Drosophila} as exceptional conditions, and the histograms of the relative positions between flies.}  The positional information of flies from the movies are shown in the three dimensional kymograph, where the boundary of the fly images are drawn in false color depending on time.  The positions of the flies are indicated as short open contours in the same color and connected by lines as visual aid. (a) Two flies moved in proximity to each other and then moved apart, causing the mask and contours to merge and split again.  (b) A missing fly image was caused by the a jumping event, followed by re-orientation of the head direction. (c)-(h) The distribution of the relative position between two flies is shown as the logarithm of the count.   The displacement is measured relative to the first fly on the perpendicular (x) direction and the parallel or the head direction (y). (c) Male 1 (M1) relative to female (F), (d) male 2 (M2) relative to female (F), (e) F relative to M1, (f) M2 relative to M1, (g) F relative to M2, and (h) M1 relative to M2.}
  \label{fig_flyJumpMerge}
\end{figure}

To increase the number of fly-fly encounters and test the ability of the repulsive algorithm to distinguish individual flies, we placed one female and two male flies in the arena. Both male flies spend a large amount of time attempting to court the female fly, resulting in frequent merging of the fly images. 85\% of the recording time, at least one of the male flies is within 6 mm of the female fly and 31\% of time both male flies are within this distance (Fig. \ref{fig_flyJumpMerge}(a)). Images are analyzed with our tracking algorithm with 100\% correct registration of flies and only one orientation reversal caused by a jumping event (Fig. \ref{fig_flyJumpMerge}(b)).  To demonstrate the tracking result, we calculated the position of each fly relative to one of the other two flies for all pairs.
Similar to previous analyses \cite{Branson2009}, we find that male flies tend to approach the female fly from the rear (Fig. \ref{fig_flyJumpMerge}(c),(d)), while maintaining a head direction oriented towards the female (Fig. \ref{fig_flyJumpMerge}(e),(g)).  The relative position of the two male flies uniformly distribute around each fly (Fig. \ref{fig_flyJumpMerge}(f),(h)), indicating no preferred orientation between males.

We compared the performance of our algorithm to the output of the CTRAX fly tracker \cite{Branson2009} using the 5-male assay, the one female, two male (1F2M) assay, and a high fly density movie trial with 16 flies in a 5 cm$^2$ arena (Table \ref{tab:trackerComparison}).  At high densities, flies frequently walk in contact with each other and jumping is more frequent.  In our repulsive contour model, where the number of flies is fixed,  identity swapping is the only  kind of error.  In comparison, because CTRAX allows the number of objects to vary, it has three types of tracking error: identity swap, lost objects, and spurious detection.  In addition, we compare the speed of the two tracking packages when run on an Intel i5 processor.  In the 5-male movie, where flies rarely come in close proximity to each other (3\% of the time), both trackers are able to track all the flies without any error.  At a higher fly density of 3 cm$^{-2}$, both trackers are able to distinguish individual flies, although CTRAX has a small portion of spurious detections and fly jumping becomes the primary source of identity swap error.  When the image flies constantly stay in close proximity, such as with frequent courtship attempts, our model is still able to locate the flies from the boundary contour, whereas CTRAX suffers from overlapping detections.  Moreover, we observe that the tracking quality of CTRAX is crucially sensitive to the user-specified input parameters such as the image value threshold and the Gaussian oval shape prior that requires multiple trials to optimize.  In comparison, the repulsive contour model only requires the knowledge of the spatial scale, either the fly length or width, and the number of flies.  The repulsive contour tracker is also 3-20 times faster than CTRAX.


\subsection{\textit{Myxococcus xanthus}  gliding motility}
\begin{figure}[!ht]
  \begin{center}
  \includegraphics[width=4in]{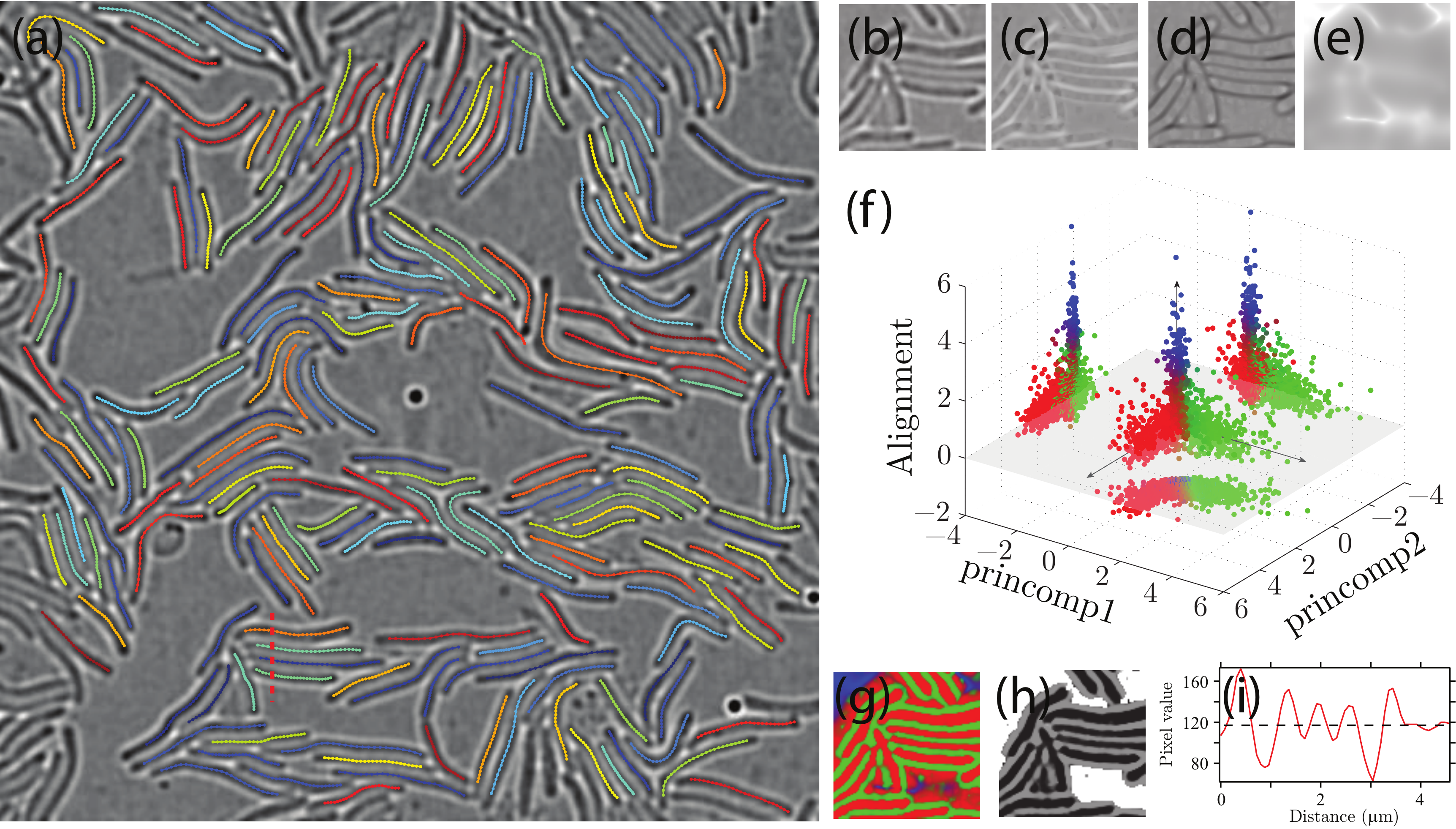}
  \end{center}
  \caption{
  {\bf Bright field images of \textit{Myxococcus xanthus} cells are taken, and transformed into the probability map that is then used to generate the image potential to interact with the contours.}  (a) A field of view of 41 $\mu$m$\times$ 41 $\mu$m (512$\times$512 pixels) contains about 200 \textit{Myxococcus xanthus} cells.  The relaxed position of the contours are overlaid on top of the bright field image with false color labeling.  (b) Zoomed-in image of a portion in (a).  (c,d) The large and the small eigenvalues of the Hessian matrix of (b).  (e) The locally averaged eigenvectors indicate the magnitude of alignment of features.  The sign is chosen such that higher value indicates less order, thus high chance to be the background and vise versa.  (f) The distribution of pixels in the classifier coordinate: two principle components from the eigenvalue-intensity space (horizontal axes), and the alignment magnitude (vertical axis).  Pixels are categorized into three groups along the three axes, and color-coded in red (ridge), green (valley) and blue (background).  The projections along three axis are shown as a guide to the eye.  (g) After classification, each pixel in the image is color coded in the same way as in (f) according to the probability of being ridge, valley or background.  (h) The enhanced image for repulsive active contour model is calculated from the classification probability map shown in (g). (i) Image intensity profile on a line segment (dashed line in (a)) illustrates the nonuniform contrast at the edge and at the inside of a cell cluster.}
  \label{fig_myxoImgProc}
\end{figure}

To highlight the ability of the repulsive contour technique to track densely packed objects, we analyzed movies of 2D swarms of \textit{Myxococcus xanthus} cells.
\textit{Myxococcus xanthus} is a soil bacterium that forms complex, 3D group structures by gliding along solid surfaces\cite{Kuner1982}.
We immobilized \textit{Myxococcus xanthus} cells between an agarose gel and a glass surface such that cells form a single layer at the interface.  Using bright field microscopy, the bacterial cell body appears dark and is surrounded by a bright halo (figure \ref{fig_myxoImgProc}(a)).  Cells are often tightly packed,resulting a poor image contrast between the neighboring cells.

Contrast-enhancement techniques such as Differential Interference Contrast (DIC) and Phase Contrast (PC) allow for the visualization of low-contrast objects but produce images whose intensity patterns are complex, usually containing both bright halos and dark regions for each object in a non-linear representation. We preprocessed the raw microscopy images before calculating the image-based potential for contour relaxation.  We first calculated the eigenvalues and eigenvectors of the Hessian of the images to quantify features such as valleys (the cell bodies) and ridges (the halos around the cells).  We observe that the two eigenvalues of the Hessian matrix and the pixel value of all pixels roughly fall in a plane in three-dimensional parameter space.  Hence, the first two principle components of these three parameters are used to specify image features.  Qualitatively, the pixels are distributed in two clusters within this two-dimensional projection (the two lobes in the projection in figure \ref{fig_myxoImgProc}(f)), with valley pixels residing in one cluster and the ridge pixels in the other, and all background pixels within the junction region of the two clusters.  
To better distinguish pixels from the background and from the high cell density area where the pixel intensity contrast is low, we consider the alignment of the features in  neighbouring areas.  Analogous to the quantification of magnetization, we treat the difference of the two eigenvalues as the magnitude of a dipole, whose direction is defined by the eigenvector associated with the larger eigenvalue. 
The locally averaged dipole moment is the order parameter that describes how ordered nearby features are aligned.  The alignment order parameter, together with the two principle components of the eigenvalue-image intensity space, forms a three-dimensional space in which the valley, ridge and background pixels are separated (figure \ref{fig_myxoImgProc}(f)).  The orientation of the three sets is determined using an expectation-maximization algorithm and the pixels are clustered into three groups according to the coordinates along the major direction of each group (Fig. \ref{fig_myxoImgProc}). We undersegment the valley and the background to prevent areas of high image potential, because the repulsive contour model is robust against undersegmented ridges but prone to errors caused by oversegmented valleys.  The enhanced image used to calculate the image potential is computed using
\begin{equation}
  I = p_{b} + 0.5p_{r},
\end{equation}
with
\begin{equation}
  p_{b} + p_{r} + p_v = 1,
\end{equation}
where $p_b$ is the probability of a pixel being the background, $p_r$ being the ridge, and $p_v$ being the valley.  $I$ is Gaussian blurred with $\sigma=1.5$ pixels before taking the gradient to smooth the image force field (Fig. \ref{fig_myxoImgProc}d).

Contours of 25 control points are evolved in an image gradient calculated from the preprocessed images (Fig. \ref{fig_myxoImgProc}d).  The repulsive force between control points on nearby cells is taken to be a quadratic function of the distance
\begin{equation}
F_{rep} =
\begin{cases}
(1-\|\mathbf x_{ij}\|/d_0)\mathbf x_{ij} & \colon  |\mathbf x_{ij}\|<d_0 \\
0&\colon |\mathbf x_{it}\|\ge d_0
\end{cases},\end{equation}
where $\mathbf x_{ij}$ is the distance vector between two control points on two different contours. $d_0$ is a cut-off distance that sets the length scale of the repulsive force.  We set $d_0$ to be 70\% of the averaged cell width, about 7 pixels.

\begin{figure}[!ht]
  \begin{center}
  \includegraphics[width=4in]{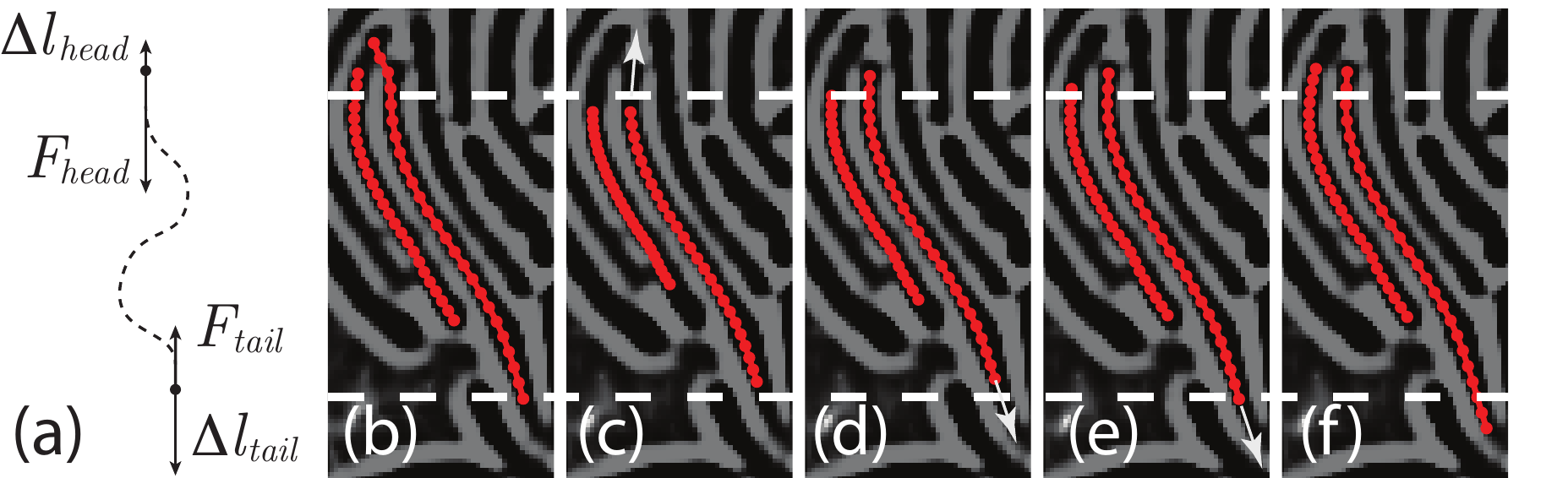}
  \end{center}
  \caption{
  {\bf A contour can ride across a ridge and cause a marginally stable configuration.  Allowing tips to grow solves this problem.} (a) The total growth of the two ends of a contour is determined by the current length and the normal length of the contour.  The growth is distributed unevenly to two termini according to the tangential resistance force.  The dashed line indicates a hypothetical contour and the two ends are indicated by the round dots.  (b) Without the ends shrinking and regrowing, the contours are trapped in a marginally stable configuration, where the contour on the right (indicated by the red dotted line) leaks to the left pocket and squeezed left contour short due to repulsion.  (c-f) The length of the contours are shortened initially and let grow to give rise to the correct contour positions.  The white arrow indicates the direction of the tip growth, and the white dashed lines are the visual guide to help illustrate the growth.}
  \label{fig_myxoTipGrowth}
\end{figure}

To prevent contours from starting the relaxation procedure at an initial position that crosses a ridge in the image (Fig. \ref{fig_myxoTipGrowth} ), we reduce the length of all contours in the first iteration of each time step to eliminate possible crossings between contours and ridges, and let the length grow back to their normal value during iterations following the image potential valleys. The length growth is implemented by adjusting the targeted contour length $l_0$ in equation \ref{eq::stretchingF} during each iteration by an amount
\begin{equation}
  \Delta l_0^\tau = \tanh(l_0 - l^{\tau-1}),
\end{equation}
where $l^\tau$ and $l_0^\tau$ are the apparent length and target length at iteration $\tau$, and $l_0$ is the normal length.  The length modification is split between the two termini of the contours according to the resistant forces exerted on the termini:
\begin{eqnarray}
 \Delta l_{head}^\tau &=& \frac{F_{tail}}{|F_{head}|+|F_{tail}|} \Delta l_0^\tau\\
 \Delta l_{tail}^\tau &=& \frac{F_{head}}{|F_{head}|+|F_{tail}|} \Delta l_0^\tau,
\end{eqnarray}
where $F_{head} > 0$ and $F_{tail}<0$ are the tangential components of the image and repulsive forces exerted on the termini.  Figure \ref{fig_myxoTipGrowth} demonstrates how this inchworm-like motion corrects misplacement of the contours.   With the length initially shortened, the faulty crossing configuration is eliminated (Fig. \ref{fig_myxoTipGrowth}b).  The upper tip grows until it hits the ridge and is unable to grow further due to steric constraints (Fig. \ref{fig_myxoTipGrowth}c, d).  The lower tip keeps growing along a narrower valley until the length reaches the target length (Fig. \ref{fig_myxoTipGrowth}e, f).

We have also included several non-critical optimizations to the dynamics of the contours for better convergence.  The tangential component of the force exerted on each control point is individually calculated and the average value is applied on each control point.  The arc distance between control points is also uniformly redistributed every 5 iterations.  These optimizations improve the relaxation in the tangential direction without qualitatively changing the behavior of the contours.

\begin{figure}[!ht]
  \begin{center}
  \includegraphics[width=4in]{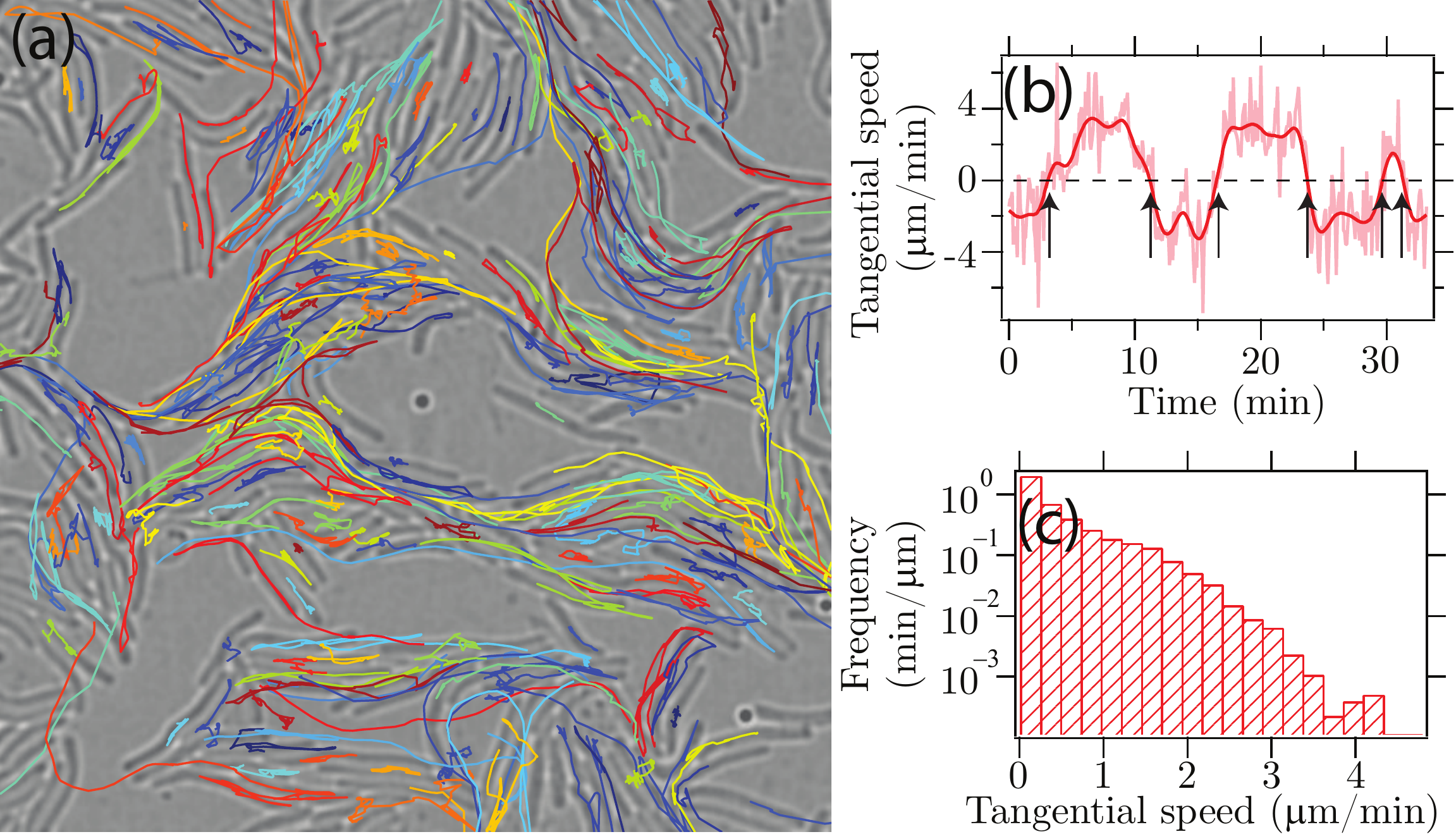}
  \end{center}
  \caption{
  {\bf Tracking results and speed statistics of \textit{Myxococcus xanthus} cells.}  (a) Trajectories of 202 cells over 2000 seconds are indicated in different colors, overlaid on the first frame of the movie. (b) The tangential gliding speed along a selected cell is plot against time.  Raw speed trace (light red) is smoothed by Gaussian kernel with $\sigma=72$ seconds (dark red).  Six directional reversals are identified by the zero-crossings of the smoothed speed curve (pointed by arrows).  (c) The histogram of gliding speed magnitude of 205 cells roughly follows an exponential distribution with the mean value at 0.49 $\mu$m/min.}
  \label{fig_myxoSpeed}
\end{figure}

We used the repulsive contour model as described above to track \textit{Myxococcus xanthus} motion within a 41 $\mu$m$\times$41 $\mu$m field of view consisting of 512$\times$512 pixels (Fig. \ref{fig_myxoImgProc}a).  205 cells are initially located in the field and we tracked these cells for 400 frames at 12 frames/min.  During the time course of the movie, cells typically move a distance about equal to 10 times their cell length and the local adjacency order is completely altered for most of the cells.  Our algorithm successfully tracks 204 cells of the total 205 cells with one cell leaking into a neighboring cell image that entered the field in the middle of the movie.  Figure \ref{fig_myxoSpeed}(a) shows the temporal projection of traces of all 205 cells, labeled in different colors and overlaid on the first frame of the movie.  These traces indicate that cells are capable of smoothly turning while gliding and reversing direction of motion.  The tangential speed of a selected cell with high motility shows the velocity estimation from position data, and indicates a directional reversal approximately every 7 minutes (Fig. \ref{fig_myxoSpeed}(b)).  Figure \ref{fig_myxoSpeed}(c) shows a histogram of the tangential speed of all 205 tracked cells while remaining in the field of view.  The speed roughly follows an exponential distribution with a mean value at $0.49 \mu$m/min.  This value is significantly smaller than previously reported results\cite{Spormann1995, Sun2011}, because we included the cells that are not motile at all, and the Gaussian smoothing altered the shape of the speed distribution.  The high cell density also reduces the mean speed as shown in \cite{Spormann1995}. 

\section{Discussions}\label{discussions}
Active contour models are widely applied in detecting features with high contrast such as boundaries in an image and is shown to be successful especially for closed contours.  Several studies applied the concept of active contour in detecting open contours with finite length, for instance actin filaments.  On the other hand, along with other machine vision techniques that segment targets from an image, active contours can be applied on individual images in time-lapsed movies of particle tracking problems.  The solution is usually separated into two stages: the objects or particles are first identified in each frame individually, then the correspondence is assigned between objects in different frames.  Merging/disappearing or splitting/emerging are allowed with specific statistical properties, usually in a maximum-likelihood fashion based on the pixel intensity level, shape changes or moving distance.  In these two-stage approaches, the object assignment is limited by the object detection quality and it is not  straight-forward to use the inter-frame object assignment information to assist particle detection.  Optimizing assignments across multiple frames can deal with single frame tracking error but dramatically increases computational cost.  In our approach, we treat the temporal dimension as an extrusion of the spatial dimension of the images and the object trajectories as elastic continua that interact with both the  image time-series and  each other in terms of repulsive forces.  Similar to the penalty based on moving distance in the linear assignment problem, the elastic energy in the temporal dimension penalizes trajectories that deviate from linear motion.  One difference between this method and the two-stage solutions is that the object identity correspondence assignment is combined with the spatial localization using image intensities.

As a closely related method to the active contour but with certain advantages, a level set method is insensitive to the topology of the contours, and therefore ideal in dealing with contours with unspecified topological features.  One drawback that limits the application of active contour and level set methods is the difficulty in applying shape constrains to the contours.  This problem can be partly solved by using a shape prior in the level set approaches at the cost of computational time.  However in many applications, the characteristic topological and geometrical features remain conserved through the entire movie, for instance the fly shape and size in the first example, cell length in the second example and the number of objects (flies, cells) in the both examples above.  These conserved geometric quantities can be predetermined and used as constraints on the solution that effectively acts  as an explicit shape prior but with a simpler form.  In addition to resembling the physical shapes with explicit lengths of the contours, our model also captures the mutually exclusive nature of these objects by adding a repulsive force between objects. This effectively prevents trajectories from crossing or collapsing into one image potential minima.

Another drawback of active contour methods is the typical sensitivity to the initial position and local minima in the image potential.  In the second example, ambiguous segmentation of cells takes place frequently due to the poor quality of the raw images, causing faulty gaps between cells and bumps along the cell body, especially in area of high cell density.  Upon sudden cell acceleration,  linear prediction may place the initial position into a faulty local minimum as shown in figure \ref{fig_myxoTipGrowth}(b).  Among various segmentation algorithms, the watershed is particularly suitable in dealing with varying boundary values, at the cost of the convenience to control the geometric properties of the segmented area.  We adopt the idea of a watershed algorithm in our active contour model in terms of dynamically varying the contour length, where the tip growth is governed by the resistant forces.  This tip growth scheme is directly analogous to the case of one dimensional watershed method, where the expansion of the two boundary points is inversely proportional to the steepness of the potential well.  The combination of active contours and tip growth is essentially a watershed algorithm with an explicit shape prior that has the capability to correct mistakes caused by improper initial positions (figure \ref{fig_myxoTipGrowth}(c-f)).

Image preprocessing can be  beneficial, sometimes essential, in order for the active contours to relax into the correct potential minima.  Because efficient dynamics of the contours requires a smooth potential landscape with slowly varying gradients, we use Gaussian smoothing before taking the gradient, or equivalently convolved the images with a derivative of gaussian (DoG) kernel.  Several factors determine the blurring radius, including the spatial scale of the objects (fly or cell width in the examples shown) and the distance between the predicted and the true position of the objects, usually at the same scale of the moving distance between frames.  The blurring kernel can be set variable depending on the motion of the objects, such as in the fly tracking case, where the fly undergoes normal walking behavior and occasional large distance jumps.  We use a Gaussian kernel with the radius the same as the fly width for the former case to achieve fast performance, and if a jump is detected, a global linear attractive potential that is only visible to the jumping flies is added on top of the original kernel for these jumping flies.  The image preprocessing is more essential and subtle in the second example.  Comparing to the raw images, the pixel intensity at the cell body, cell boundary and background is more uniform after we convert the pixel intensity into assignment probability of three pixel categories.  The elevated background intensity prevents contours from leaking to the empty background, which is a severe problem if the image is not preprocessed (data not shown).  Pixel classification also enhances the barrier between cells and normalizes the pixel values on the cells.

Since the evolution of the contours follows explicit dynamics indicated in equation \ref{eq::discretedynamics}, it is convenient to implement additional factors that affect the tracking results in terms of explicit forces, or to modify the elastic property to fit particular shape requirements.  The length constraints and tip growth are applied in the form of explicit forces in the case of tracking flies and \textit{Myxococcus xanthus} cells.  In the first example, we also add energy costs to penalize sharp turns, which is observed to cause identity swap in rare ambiguous frames.  In addition to applying explicit forces, the internal degrees of freedom of the contour curvature allows extra flexibility in modeling more complex shapes.  For instance, the bending elasticity can be set as a variable along the contour instead of a constant,  allowing parts of the contour to act as soft joints and the rest as hard stems.  This is especially suitable in tracking objects with internal motion such as head turning in mice.

Because the image information in adjacent frames is considered as a whole in the localization, minor corruptions in image quality can be corrected by the adjacent frames.  On the other hand, major corruption can cause propagating errors in the following frames.  The relaxation to the right image potential minimum suffers especially when the image potential energy landscape is highly curved with a large number of local minima, such as in the case of \textit{Myxococcus xanthus} images where cells are bent and intertwined with identical local statistical character of pixel intensities at cell width length scale.  Inchworm tip growth has the ability to correct initial misplacement given that the cells only moves tangentially by a  small amount (20\%) relative to the cell length within two frames.  Our algorithm also does not consider the case of new objects entering the image.  New objects can be detected using other methods such as level set and then treated as a regular repulsive active contour.

Comparing our active contour tracking model that uses the explicit dynamics of the contours with previous Bayesian approaches such as Branson et al. \cite{Branson2009}, the essential principles are correspondingly similar to models of the physical properties of the objects.  Active contour models treat images as a potential energy landscape, and the objects are described by a set of contours with specific lengths and bending properties to define the shape.  Contours move according to the image potential with certain damping factor until relaxing at a minimum, and we add repulsion forces between objects to prevent merging.  Correspondingly in Bayesian approaches, each image is treated as a spatial distribution of pixel vales, and an object is a parameterized distribution model with particular shape priors, such as the covariance and centroid of a Gaussian distribution.  Tracking   maximizes the posterior likelihood of the modeled distribution explaining the image, which involves a particular numerical gradient descending scheme similar to the damped dynamics of contours.  A Gaussian mixture model is adopted for the case of multiple object tracking, which clusters pixels into groups with small or no overlaps resembling repulsion between Gaussian mixtures.

\section{Conclusions}\label{conclusions}
We developed a multiple-object tracking algorithm for time-lapsed movie based on active contour model, taking into account the physical exclusion of the objects.  In our model, multiple oval-shaped or elongated curved objects are represented by open elastic contours with individually fixed length in each frame, and the motion of objects in time is treated as an extrusion in the temporal dimension, thus the spatio-temporal kymograph are modeled as mutually repulsive elastic membranes.  We illustrated the application of the repulsive active membrane model on two sets of realistic experimental data, tracking multiple \textit{Drosophila} walking and chasing, and tracking individual curved gliding bacteria \textit{Myxococcus xanthus} at high density.  Individual objects are successfully tracked at high efficiency at video frequency with low error rate ($10^{-4}$) that can be conveniently corrected in separate steps.

\section*{Acknowledgments}
We sincerely thank Mala Murthy for providing support in data acquisition and helpful discussions.  We also acknowledge funding from the Pew Charitable Trusts and the National Science Foundation.

\bibliography{YDthesis}

\newpage
\section*{Tables}

\begin{table}[!ht]
\caption{
\bf{Comparison between CTRAX and the repulsive contour model.} For all three movies tested, the number of fly pairs that have a distance closer than twice the fly width is counted as a fly-fly contact. In speed comparisons, both trackers ran on the same single CPU core on the same platform.}
\begin{tabular}{|c|c|c|c|c|c|c|}
\hline
\multirow{2}{*}{Assay} & Length of Movie & Fly-fly & \multicolumn{2}{c|}{\textbf{CTRAX}} & \multicolumn{2}{c|}{\textbf{Repulsive Contours}}\\
\cline{4-7}
& (frames) & contacts & Error & Speed (fps) & Error & Speed (fps)\\
\hline
5 males & 20203 & 1427 & 0 &4.4& 0&40 \\
\hline
1F2M & 9000 & 12956 & 25 &3.5 & 0&70\\
\hline
16 flies &4581 & 8717 & 9&3.1&3&10\\
\hline
\end{tabular}
\label{tab:trackerComparison}
\end{table}

\newpage

\section*{Figures}

\end{document}